# Surface Plasmon Enhanced Strong Exciton-Photon Coupling in Hybrid Inorganic-Organic Perovskites Nanowires

Qiuyu Shang[1,3], Shuai Zhang[3], Jie Chen[1,3], Pengfei Yang[1], Chun Li[1], Wei Li[4], Yanfeng Zhang[1], Qihua Xiong[5*], Xinfeng Liu[3*], Qing Zhang[1,2*]

[1]Department of Materials Science and Engineering, College of Engineering, Peking University, Beijing 100871, P. R. China

[2]Research Center for Wide Gap Semiconductor, Peking University, Beijing 100871, China

[3]CAS Key Laboratory of Standardization and Measurement for Nanotechnology, CAS Center of Excellence for Nanoscience, National Center for Nanoscience and Technology, Beijing 100190, P. R. China

[4]Department of Physics, Tsinghua University, Beijing 100084, P. R. China

[5]Division of Physics and Applied Physics, School of Physical and Mathematical Sciences, Nanyang Technological University, Singapore 637371

Corresponding[*]: Q_zhang@pku.edu.cn; Liuxf@nanoctr.cn; Qihua@ntu.edu.sg

**Abstract:**

Manipulating strong light-matter interaction in semiconductor microcavities is crucial for developing high-performance exciton polariton devices with great potentials in next-generation all-solid state quantum technologies. In this work, we report surface plasmon enhanced strong exciton-photon interaction in $CH_3NH_3PbBr_3$ perovskite nanowires. Characteristic anti-crossing behaviors, indicating Rabi splitting energy up to ~ 560 meV, are observed near exciton resonance in hybrid semiconductor-insulator-metallic waveguide cavity at room temperature. An exciton-photon coupling strength enhancement factor of ~ 1.4 times is evaluated, which is mainly attributed to surface plasmon induced localized excitation field redistribution. Further, systematic studies on nanowires and insulator dimension dependence of exciton-photon interaction are presented. These results provide new avenues to achieve extremely high coupling strengths and push forward the development of electrically pumped and ultra-low threshold small lasers.

**Introduction:**

Strong light-matter interaction in solid state systems, distinguished from weak light-matter interaction by arising of a new quantum state with half-light and half-matter, is particularly interesting for novel photonic and quantum devices. In semiconductor cavities, exciton polaritons are formed when the interaction rate $g$ between exciton and cavity photon is faster than their dissipation rate $\gamma$, so called "strong coupling regime".[1] Being a Bosonic quasi-particle with effective mass four orders lower than that of free electrons,[2] exciton polariton not only provides an ideal platform for macroscopic quantum electrodynamics, but also holds great promise for realizing superfluities,[3, 4] Bose-Einstein condensation (BEC) at room temperature [5-10] and thereby ultra-low threshold lasing [11-13] and slowing light devices.[14, 15]

Pioneer works have been made on strong coupling between microcavity photons and excitons in materials systems with intrinsic oscillation strength such as organic semiconductor,[11] wide bandgap semiconductors (GaN, ZnO),[12, 13] two dimensional semiconductor [16] or enhanced oscillation strength by quantum well structures (GaAs, CdTe, *etc*).[9, 10] However, most of these polariton effects are achieved under either cryogenic temperature, ultra-violet regime, or relying on expensive and complicated cavity fabrication proceedings. Hybrid inorganic-organic lead halide perovskites, combing advantages of both organic and inorganic materials, may provide a great platform to realize versatile, low cost exciton-polariton and polariton lasing.[17-21] Perovskites exhibit facile engineering of exciton properties throughout ultra-violet to near-infrared regime. Their large intrinsic exciton oscillation strength effectively relives the need of high factor (Q) cavity for polariton device over the whole visible spectra regime.[18, 22] Further, the great electrical properties extremely raise the possibility of realizing electrically driven polariton devices. With two end-facets as reflector, perovskite nanowires (NWs) naturally form active optical microcavities, which have shown impressive lasing properties, such as high coherence and ultra-low threshold.[23-26] Recently, we demonstrate that single crystalline perovskite NWs ($CH_3NH_3PbBr_3$, $MAPbBr_3$) support strong room temperature polariton effect in visible range.[27] Although the Rabi splitting energy (~ 390 meV) is higher than those reported in similar semiconductor microcavity (< 200 meV),[28, 29] the lasing of as-grown $MAPbBr_3$ NW is still govern by population inversion of excitons. Considering the increasing scattering processes at intense pumping condition, it is critically important to explore novel methods to promote strong exciton-photon coupling strength to maintain coherence of excitons and photons for the practical application of polariton.

Exciton-photon coupling strength can be described as, $g \propto \sqrt{n*f/V_{eff}}$, where $n$, $f$ and $V_{eff}$ are oscillator numbers per volume, exciton oscillator strength and effective mode volume, respectively. According to the equation, exciton-photon coupling frequency $g$ can be enhanced via (1) enlarging oscillator numbers by improving quality factor $Q$ of optical cavities or increasing the amount of active materials; (2) promoting oscillator strength by, *i.e.*, adopting quantum confinement structures; (3) lowering mode volume by pushing optical cavities dimension into sub-wavelength regime.[1, 22] For example, for technologically important semiconductor GaAs with small intrinsic oscillator strength, room temperature exciton polariton can be realized by coupling GaAs quantum wells structure to high-$Q$ Bragg reflector cavity.[7-9] The need for high-$Q$ cavities and quantum well structures can be partly mitigated by adopting reduced dimensional cavities to confined photonic modes, *i.e.*, one-dimensional NW of semiconductors.[22] Surface plasmon owns unprecedented capability of sub-wavelength mode confinement and strong localized field generation. It has been used to tailor light matter interaction in weak coupling regime.[30-35] As a contrast, the reports on surface plasmon polariton tuned strong exciton-photon coupling effect are still inadequate.

Herein we introduce metal-insulator-semiconductor (MIS) hybrid plasmonic cavity to enhance strong exciton-photon coupling strength of lead halide perovskites. Rabi splitting energy near exciton resonance of MAPbBr$_3$ NWs is enhanced by ~ 1.4 times (up to ~ 560 meV) through coupling the NW with silica/silver film, which is mainly due to strong localized excitation field and reduction of effective mode volume $V_{eff}$ induced by surface plasmon. Moreover, the enhancement factor increases with the decreasing of silica thickness. Further, we prove that Exciton-photon coupling strengths of both perovskite and hybrid perovskite-metallic nanowires are enhanced with the decreasing of nanowire dimension due to reduction of photonic modes. These results are helpful for promoting exciton-photon coupling towards development of continuous wave, electrically driven polariton lasers, slow light, coherence light sources and nonlinear optics.

## Results and Discussions

**Figure 1a** shows the schematic diagram of strong coupling between excitons and photons in semiconductor NWs. With two end facets serving as micro-reflectors, semiconductor NW is naturally functioned as a Fabry-Pérot optical microcavity. Cavity exciton polaritons form if coupling frequency *g* between exciton and cavity photon is faster than all decoherence processes in the microcavity, leading to the hallmark anti-crossing features near to exciton resonance in absorption or reflectance spectroscopy. In NW microcavity, photon confinement arises, leading to a reduced $V_{eff}$ due to the large refractive index contrast between semiconductor NWs and its surroundings (air, glass, silver), [36] which consequently leads to enhancement of exciton-photon coupling strength. **Figure 1b** shows the schematic diagram of as-proposed hybrid plasmonic microcavity, which consists of single crystalline MAPbBr$_3$ NW, a silver (Ag) film and a thin protective layer SiO$_2$ between MAPbBr$_3$ NW and silver film. When the NW is placed on an insulator-silver film substrate (**Figure 1b**), hybrid surface plasmon polariton (SPP) modes arise and strongly confines electromagnetic modes into sub-wavelength volume around the nanoscale silica layer (**Figure 1d**). Effective mode volume in NWs is reduced by the strong confinement. On the other hand, the localized SPP field also serves as a vacuum field *E* and redistributes the single photon excitations, which leads to great enhancement of oscillator numbers density *n* which is proportional to the electric field intensity $|E|^2$. With the decrease of $V_{eff}$ and increase of *n*, the strong exciton-photon coupling strength *g* is enhanced accordingly.

**Figure 1c** show scanning electron microscopy (SEM) of a typical perovskite NW. The as-grown MAPbBr$_3$ perovskite NW exhibits uniform rectangle facets which are expected to atomically smooth (**Figure 1c, inset**, scale bar: 0.5 μm). The length and width of as-grown perovskites are ~ 200-700 nm and ~ 2.5-20 μm, respectively. To focus on interaction between exciton and photonic mode, the NWs with width of ~ 400-600 nm are selected. In this dimension, plasmonic mode has much higher propagation loss than photonic mode due to Ohmic absorption. On the other hand, in order to reduce scattering and radiation loss, growth proceedings of Ag film (thickness: ~ 50 nm) and SiO$_2$ layer (thickness: ~ 5-20 nm) are optimized to obtain highly smooth surface (root mean surface roughness: ~ 1 nm, **Figure S2a**).

**Exciton-Polariton in Perovskite NWs on Glass and SiO$_2$/Ag Substrates**

Next, we probe room temperature exciton-polariton properties in individual perovskite NWs on glass (**Figure 2a-c**) and 5 nm SiO$_2$/Ag substrates (**Figure 2d-e**), respectively, by a home-built *in-situ* excitation and remote collection confocal micro-spectroscopy system. As shown in **Figure 2a and 2d**, a focused laser (405 nm, continuous wave, spot size: ~ 1 μm) excites one end of a NW while emission is detected at the other end. This measurement system gives a clear observation of longitudinal standing wave modes along NW microcavities. To exclude size effects, NWs with similar width (~ 550±30 nm) and length (~ 8.4±0.5 μm) are selected. *In-situ* photoluminescence (PL) spectroscopy of perovskite NW shows single symmetric peak due to exciton recombination for both substrates ($P_1$, green solid lines;

glass, **Figure 2b**; 5 nm SiO$_2$/Ag, **Figure 2e**). The PL peak center of the NW on 5 nm SiO$_2$/Ag substrate is red-shifted by ~ 6 nm compared with that on glass substrate (~ 537 nm). The redshifts may be due to Franz-Keldysh effects, which will be discussed further.[37] Further, the full with at half maximum (FWHM) of PL peaks are ~ 20 nm and ~ 23 nm for NWs sitting on glass and 5 nm SiO$_2$/Ag substrates, respectively. The corresponding damping constants are $\gamma_{ex}$ ~ *c. a.* 65 meV and 80 meV, which agree well with the value measured by reflection spectroscopy (59 and 77 meV, **Figure S4d**).[38] The increase of exciton damping on SiO$_2$/Ag substrate is attributed to introducing of nonradiative exciton-plasmon scattering process. Compared with the *in-situ* emission, remote collected PL spectroscopy show significant redshift ($P_2$, navy solid lines), which is primarily attributed to self-absorption (**Figure S3a**).[39] Moreover, a series of narrow peaks are clearly resolved above exciton emission background. The narrow peaks can be well fitted by Lorentzian line-shapes with FWHM of ~ 15 meV. The spacing between adjacent oscillation modes, or so called free spectra range (FSR), is inversely proportional to NW length $L_z$, suggesting that the oscillation modes are assigned to longitudinal Fabry-Pérot resonant modes along the NW microcavities (**Figure S3b**).

To explore exciton-photon coupling in MAPbBr$_3$ NWs, energy (*E*)-wavevector (*k*) dispersions of microcavity polaritons are extracted out, as shown in **Figure 2c** (glass) and **Figure 2f** (5 nm SiO$_2$/Ag) respectively. The detailed method can be found in **Methods.** The dispersion curves of cavity photons (navy solid line) and excitons (black dashed line) are also presented as a comparison. In both configurations, *E-k* dispersion can be well fitted based on a two-coupled oscillator model (navy solid curve) (**Supplementary Information, Note 1**), which clearly exhibits an anti-crossing at the exciton resonance due to the quantum-mechanical noncrossing rule for two interacting energy levels.[22] No emission from upper polariton branch is detected due to relatively larger optical losses.[40] The calculated Rabi splitting energy, evaluated by minimum vertical distance between lower and upper polariton branches are ~ 270 and 387 meV for NWs on glass and 5 nm $SiO_2$/Ag substrates, respectively. In both configurations, the measured Rabi splitting energy is larger than photon and exciton escaping rate ($\gamma_{ph}$ and $\gamma_{ex}$: ~ 15 and 59 meV for NW on glass substrate; ~ 16 and 77 meV for NW on 5 nm $SiO_2$/Ag substrate). It suggests that the condition of strong exciton-photon coupling is satisfied.

Further, the Rabi splitting energy for NW on 5 nm $SiO_2$/Ag substrate is 1.43 times of NW on glass substrate. Two reasons could be responsible for the enhancement of coupling strength: 1) reduction of effective mode volume and 2) local field enhancement. To explore the underline physics of the coupling enhancement, mode simulation and electric field distribution are calculated for NWs on glass and 5 nm $SiO_2$/Ag substrates (**Materials and Methods**), as shown in **Figure 3**. In glass configuration (**Figure S6e-h**), photonic modes including HE11-like, EH11-like, TE01-like and TM01-like modes are supported when the nanowire width is 400-600 nm. Although all of these photonic modes can couple to excitons, only the fundamental HE11-like mode owning highest effective refractive index and relatively lower loss is dominated. In the other words, the coupling between excitons and the other photonic modes can be eliminated. The upper panel in **Figure 3a** shows a two-dimensional projection (Y-Z plane) of the HE11 mode in a 550 nm × 8.4 μm nanowire at $\lambda$ = 550 nm, which propagates along the long axis of NWs. In SiO$_2$/Ag configuration, hybrid SPP modes arise together with the photonic modes (**Figure S6i-l**). The lower panel in **Figure 3a** shows a two-dimensional projection of the plasmonic mode, which has ultra-small effective mode volume (**Figure S6i**). nevertheless, the huge dissipative rate due to intrinsic metallic Ohmic loss could not be easily overcome for SPP mode.[41] As a contrast, HE11 photonic mode with low loss can still strongly couple to excitons in MIS configuration. The domination role of HE11 photonic mode coupling with excitons is confirmed by polarization dependence emission of NW in MIS configuration (**Figure S7c**). Later, we applied the HE11 mode to calculate the effective mode volume $V_{eff}$ in both configurations (left: glass, right: 5 nm SiO$_2$/Ag), as shown in **Figure 3b**. The calculated effective mode volume $V_{eff}$ (**Materials and Methods**) of HE11 mode of NWs is 1.01 times of physical volume in 5 nm SiO$_2$/Ag configuration while 1.09 times in glass configuration. The HE11 mode volume in MIS configuration is slightly smaller that of photonic configuration by 7%. Further, the electric field of MAPbBr$_3$ NWs is re-distributed with the present of Ag film, as shown in **Figure 3c**. On glass substrate, electric field is confined symmetrically inside the MAPbBr$_3$ NWs; while on 5 nm SiO$_2$/Ag substrate, the electric field is no longer symmetric around NWs but strongly confined near to perovskite-SiO$_2$-Ag interface. On 5 nm SiO$_2$/Ag substrate, the maximum localized electric field which dominates the enhancement increases by 1.75 times according to the simulation.[42, 43] Eventually, as $g \propto \sqrt{n*f/V_{eff}}$, the overall strong coupling strength of MAPbBr$_3$ NW on 5 nm SiO$_2$/Ag substrate is nearly 1.4 times that of bare NWs on glass substrate, which agrees well with experiment values (1.43).

Further, thickness of SiO$_2$ layer is varied to tune SPP field and therefore exciton-photon coupling in microcavities. **Figure 4a** includes *E-k* dispersion of MAPbBr$_3$ NWs in MIS configuration with SiO$_2$ thickness from 5 nm (navy curve), 10 nm (pink curve) to 20 nm (orange curve) respectively. The widths and lengths of the NWs are ~ 550±30 nm and ~ 8.4±0.5 μm. Rabi splitting energy increases from ~ 270 meV (glass) to ~ 387 meV (5 nm SiO$_2$/Ag) respectively. The detailed PL spectroscopy can be seen in **Figure S5.** Also, The overall dispersion curve moves towards $k_{//}$=0 point with the decreases of SiO$_2$ thickness. Reflection spectroscopy and spectroscopic ellipsometer are conducted to determine the damping constant $\gamma$ and background dielectric constant $\varepsilon_b$ (**Materials and Methods**). With the decreasing of SiO$_2$ layer thickness, the exciton damping loss $\gamma$ gradually increases due to stronger exciton-plasmon interaction. The background dielectric constant $\varepsilon_b$ deceases from 4.7 to 3.3, which steepens photon mode in microcavities and therefore push low polariton branch towards zero-points (**Figure S4**). Mode and field simulation are further conducted to explore the SiO$_2$ thickness dependence of coupling strength. The effective mode volume of HE11 mode shows little difference when SiO$_2$ thickness varies (**Figure 3b**). As a contrast, electric field inside MAPbBr$_3$ NWs are extensively redistributed by SPP mode. The maximum field intensity which decreases gradually as SiO$_2$ thickness increases from 10 to 100 nm (**Figure 4c** insets). When the thickness approaches ~ 100 nm, Rabi splitting energy of NWs in MIS configuration is nearly as the same as NWs on glass substrate. The $SiO_2$ thickness dependence of Rabi splitting energy agrees well with the variation of localized electric field amplitude. It further confirms that enhancement of exciton-photon coupling strength is mostly due to redistribution of electric field by surface plasmon.

Time-resolved photoluminescence spectroscopy (TRPL) of $MAPbBr_3$ NWs on different substrates is conducted to verify SPP enhanced local field. As shown in **Figure 4d**, average PL lifetime of the $MAPbBr_3$ NWs with the widths and lengths being ~ 550±30 nm and ~ 8.4±0.5 μm shortens gradually from 1.1 ns, 0.8 ns, 0.2 ns to 0.1 ns when the thickness of $SiO_2$ decreases. According to Purcell effect in weak coupling regime, the radiative transition rate of emitter is modified by the density of states, *i.e.*, Purcell factor is proportional to electric field enhancement.[44] The lifetime of NW on glass substrate is 11, 5.5 and 1.4 times that of NWs on $SiO_2/Ag$ substrates when the $SiO_2$ thickness is 5 nm, 10 nm and 20 nm, respectively. The values are slightly larger than Purcell factor calculated using electric field (**Figure 4c**), which may be due to increased probability of nonradiative damping accomplished by surface plasmon.

In NW microcavity, exciton-polariton effect is highly dependent on the dimension of NWs. To further investigate surface plasmon enhanced strong exciton-photon coupling, group of NWs on glass and 5nm $SiO_2/Ag$ substrates are selected with widths ranging from ~ 400-600 nm and lengths ranging from ~ 2.5-21 μm. **Figure 5a-c** shows *E-k* dispersions of NWs with comparable width of ~ 550±30 nm and variable length $L_z$ of ~ 2.5±0.5 μm (a), 8.4±0.5 μm (b) and 20.8±0.5 μm (c), respectively. Anti-crossing features are observed for the three groups of NWs on glass and 5 nm SiO$_2$/Ag substrates. When the NW length decreases from 20.8 to 2.5 μm, Rabi splitting energy increases from 253 meV (glass)/354 meV (5 nm SiO$_2$/Ag) to 390 meV (glass)/560 meV (5 nm SiO$_2$/Ag), respectively. **Figure 5d** shows the relation of Rabi splitting energy of microcavity polariton versus the effective mode volume. When effective mode volume in NWs is pretty large ($V_{\text{eff}} > 6.05 \times 10^{-18} \varepsilon_b$ @550 nm), $V_{\text{eff}}$ is nearly as the same as the physical volume of NW $V_0$ (**Figure 5d**, pompadour color area). The exciton polariton is bulk polariton showing Rabi splitting energy $\Omega = \sqrt{2E_0(E_L - E_T)}$, with $E_L$ the longitudinal resonance energy, $E_T$ the transverse resonance energy and $E_0$ the exciton resonance energy, which remains a constant (glass: ~ 253 meV; 5 nm SiO$_2$/Ag: ~ 354 meV). However, as $V_{\text{eff}} < 6.05 \times 10^{-18}$ m$^3$, the NWs enter microcavities region with $V_{\text{eff}} > V_0$; thereby exciton polaritons become microcavity polaritons (**Figure 5d**, lavender color area) (**SI note 1**).[45-47] Rabi splitting energy is enhanced in microcavity with the reduction of NW dimensions. The dashed curves in **Figure 5d** give Rabi splitting energy calculated from the above formulas for NWs on 5 nm SiO$_2$/Ag and glass substrate. The experiment data of Rabi splitting energy extracted out from the *E-k* dispersions shows good consistent with theoretical value in both bulk and microcavity regimes.

Exciton-polariton effect in microcavities results in large dispersion and high refractive index near the excitonic resonance, which is important for optical nonlinearity, switches and storages, *etc*.[14, 15] **Figure 5e** plots group refractive index of HE11 mode of NWs sitting on glass and 5 nm SiO$_2$/Ag substrates. The energy dependence curve of group refractive index is extracted from *E-k* dispersion curves (**Figure 5a, b**) by using the classic formula $n_g = (dE/dk)_{vacuum}/(dE/dk)_{cavity}$ (**SI Note 2**). The data points on curves are directly obtained through calculating group index using *E* and *k* of each oscillation peaks in emission spectroscopy. The Strong curvature of the dispersion indicates that the group index increases significantly near exciton resonant energy, such as from 5.1 at 2.07 eV to 16.2 at 2.24 eV for the 2.5 μm-long NW on 5 nm SiO$_2$/Ag substrate. With the increasing of NW length, the coupling strength and therefore group index reduces due to the increase of mode volume. For the same reason, the group refractive index of NW is enhanced by surface plasmon, *i.e.* $n_g$ = 16.2 and 9.6 at 2.24 eV for the 2.5 μm-long NW sitting on 5 nm SiO$_2$/Ag substrate and glass substrate, respectively. It shows that the group refractive index of NW on 5 nm SiO$_2$/Ag substrate is nearly 2 times of NW on glass substrate and 8 times that of bulk regime.

**Conclusion**

In this work, strong coupling strength of ~ 560 meV is demonstrated in facile solution processed MAPbBr$_3$ NWs. The as-reported value is so far the highest in the spectra range of ~ 540 nm to our best knowledge. The results are due to three reasons: 1) MAPbBr$_3$ has intrinsic high exciton oscillation strength; 2) NW is functioned as an active microcavity which provides good photonic mode confinement; 3) surface plasmon induced EM field serves as vacuum field, redistributes oscillators and then promotes density of oscillators in effective area. Although exciton damping rate increases due to surface plasmon, the strong exciton photon coupling strength is still extensively enhanced by 1.4 times due to the three advantages. In smaller NWs, strong exciton-plasmon interaction would be expected. Moreover, the enhanced coupling strength also reduces group velocity and thus slows light speed, which is useful for slow-light applications in nonlinear optics and quantum optics. These results accelerate the advancements of room temperature ultra-low threshold polariton lasers and nonlinear devices.

**Acknowledgements**

Q.Z. acknowledges funding support from the Ministry of Science and Technology (2017YFA0205700; 2017YFA0304600) and Natural Science Foundation of China (No. 61774003). Q.Z. also thanks the support of start-up funding from Peking University, one-thousand talent programs from Chinese government, open research fund program of the state key laboratory of low-dimensional quantum physics. X.F.L thanks the support from the Ministry of Science and Technology (No. 2016YFA0200700 and 2017YFA0205004), National Natural Science Foundation of China (No. 21673054), Key Research Program of Frontier Science, CAS (No. QYZDB-SSW-SYS031).

**Author contributions**

Q.Z. conceived idea and designed experiments. Q.S., S.Z, J.C prepared the samples and performed PL spectroscopy measurements and SEM. P.Y performed AFM measurements. Q.S and J.C conducted time-resolved PL measurement. Q.S, Q.Z performed the simulations. All the authors discuss the results and the manuscript. Q.Z led the project.

**Additional information**

Supplementary Information accompanies this paper at online version of this paper.

**Competing financial interest:**

The authors declare no competing financial interest.

**Materials and Methods**

**MAPbBr$_3$ NW synthesis**. The CH$_3$NH$_3$PbBr$_3$ NWs were synthesized by a modified solution based on self-assembly method reported by *Wang et al*.[48] 105 mg PbBr$_2$ and 345 mg CH$_3$NH$_3$Br were mixed with 10 ml DMF in a 20 ml beaker to prepare MAPbBr$_3$/DMF solution precursor. The solution mixture was stirred over night at 60°C, percolated by a syringe filter (Whatman, 0.45 μm) and diluted to 0.025 mol/L. An upturned 50 ml beaker is placed into a 250-ml beaker containing 75 ml dichloromethane (DCM). A petri dish with diameter of 7.5 cm with glass and SiO$_2$/Ag substrates (size: 10×10 mm$^2$) locates on top of the 50-ml beaker. The substrates are wiped by hydrophobic polyethylene in advance. 15 μl MAPbBr$_3$/DMF solution is dropped onto the center of the substrates, which will be uniformly dispersed. The whole experimental facility is sealed by a piece of tinfoil for 24 h, followed by collection of as-prepared NWs on substrates. Ag film is prepared using a thermal evaporator. SiO$_2$ layers with different thicknesses (5, 10 and 20 nm) are deposited onto Ag film by magnetron sputtering. The surface morphology of the Ag film after deposition of SiO$_2$ is measured by an atomic force microscope (AFM) (**Figure S2a**).

**Spectrally and Spatially Resolved Micro-Photoluminescence Measurements.** For steady-state PL spectroscopy, a continuous-wave laser (wavelength: 405 nm) is focused onto an individual MAPbBr$_3$ NW using an Olympus BX51 microscope equipped with a 100× objective (NA = 0.95; spot diameter: 1 μm). PL emission signal from the NW was collected by the same microscope objective in a backscattering configuration and analyzed by Princeton Instrument spectrometer (PI Acton, Spectra Pro 2500i) equipped with a TE-cooled charge coupled detector camera (PIXIS-400B). A 405-nm long pass filter was used to block the excitation laser. The PL image is recorded by a cool-snap color camera equipped on Olympus BX51 microscope. The PL lifetime measurements were conducted by a time-corrected single photon counting technique (TCSPC). The excitation laser source is a frequency-doubled mode-locked Ti-sapphire oscillator laser (800 nm, repetition rate 76 MHz, pulse length 120 fs). The excitation fluence is very low to avoid heating and exciton-exciton scattering effects. A 405-nm long pass filter was used to filter out ambient light and pass light from the MAPbBr$_3$ exciton emission near 540 nm. For lasing measurements, the laser source is generated by frequency doubled from a Coherent Astrella regenerative amplifier (80 fs, 1 kHz, 800 nm) that was seeded by a Coherent Vitara-s oscillator (35 fs, 80 MHz). All experiments are carried out at room temperature and atmosphere condition.

**Measurement of Background Dielectric Constant $\varepsilon_b$ and Damping Constant $\gamma$.**

In Lorentz oscillator model, the dielectric function is given by, $\varepsilon = \varepsilon_b (1 + \frac{\omega_L^2 - \omega_T^2}{\omega_L^2 - \omega_T^2 - i\omega\gamma})$.[49,50] To obtain the *E-k* dispersion relation of exciton-polariton for NWs on SiO$_2$/Ag substrates, the background dielectric constant $\varepsilon_b$ and damping constant $\gamma$ need to be determined (**SI Note 1**). The background dielectric constant is calculated from the spectroscopic ellipsometer.[51-53] With the dielectric function $\varepsilon(E) = \varepsilon_1(E) + i\varepsilon_2(E)$ and the refractive index $N(E) = n(E) + ik(E)$ ranging over a wide photon-energy, the *n* and *k* could be obtained from the spectroscopic ellipsometer and finally the background dielectric constant $\varepsilon_b$ can be calculated as $\varepsilon_1 = n^2 - k^2$ and $\varepsilon_b$ is dielectric constant located at $\omega \to \infty$. The values of the damping constant $\gamma$ are measured by reflection spectroscopy.[54] As the experimental reflection spectrum can be expressed as $R = A \frac{|\sqrt{\varepsilon} - 1|}{|\sqrt{\varepsilon} + 1|} + C$, the damping constant $\gamma$ could be determined by fitting the first derivative of reflection spectra after substituting into the dielectric function. The results can be found in **Figure S4**. Spin-coated MAPbBr$_3$ perovskite film on SiO$_2$/Ag substrates (SiO$_2$ thickness: 5 nm, 10 nm and 20 nm) is prepared to conduct ellipsometer spectroscopy to obtain these parameters.

**Numerical Calculations**. Finite element method (MODE Solutions, *Lumerical*, inc) is conducted to simulate photonic and plasmonic waveguide mode in photonic NWs (on glass substrate) and plasmonic NWs (on SiO$_2$/Ag substrate). The results can be seen in **SI Figure S6 and SI Note 4**. Finite different time domain (FDTD) Solutions is conducted to calculate electric field distribution for photonic and plasmonic NWs.

Perfectly matched layer (PML) boundary conditions was applied to solve the Maxwell's equations, taking into account interaction with the Si substrate and the $SiO_2$ protection layer.

Since $L_z \gg L_x, L_y$, effective mode volume $V_{eff}$ of HE11 mode in the $MAPbBr_3$ NWs at exciton resonance energy is calculated using equation:

$$V_{eff} = \frac{\int_V \varepsilon(r)|E(r)|^2 d^3r}{(\int_{V_0} \varepsilon(r)|E(r)|^2 d^3r)/A_0}$$

$$= L * \frac{\int_A \varepsilon(r)|E(r)|^2 d^2r}{(\int_{A_0} \varepsilon(r)|E(r)|^2 d^2r)/A_0} \quad [1]$$

where $V(A)$ the simulation volume (area), $V_0(A_0)$ the actual NW volume and $A_0$ the cross-sectional area. In bulk regime, equation 1 returns to the NW crystal volume, transforming $\Omega = 2\hbar g = \hbar\sqrt{\frac{V_0(\omega_L^2 - \omega_T^2)}{V_{eff}(V)}}$ for bulk polaritons into $\Omega = \sqrt{2E_0 E_{LT}}$ (**SI Note 1**). Whereas at microcavities regime ($V_{eff} < 6.05 \times 10^{-18}$ m$^3$), the leaking of the EM field into the surroundings leads to the difference between the effective mode volume $V_{eff}$ and the actual volume $V_0$, especially for the NWs on $SiO_2$/Ag substrates.

**Figures and Captions**

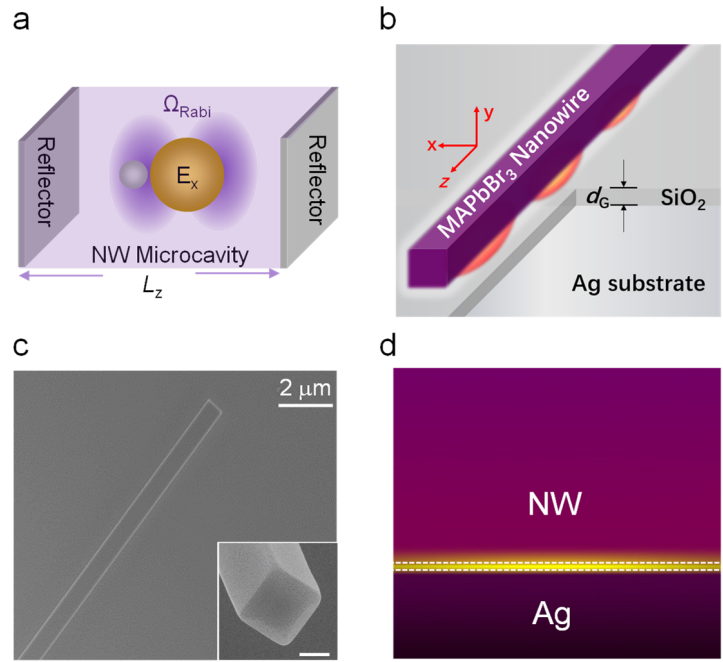

**Figure 1**. **Structure of Hybrid MAPbBr₃ Plasmonic Microcavity.** (a) Schematic of strong coupling between exciton and photon in NW microcavity. Two end facets of NW serve as micro-reflectors and therefore the NW naturally functions as Fabry-Pérot microcavity. (b) Schematic of hybrid plasmonic microcavity structure. A MAPbBr₃ NW sits on SiO₂/Ag substrate (~ 50±1 nm) with a spacer layer silica ($d_G$ ~ 5–20±1 nm). (c) Scanning electron microscopy (SEM) image of a representative MAPbBr₃ NW sitting on the Ag measurement substrate (scale bar: 2 μm; inset scale bar: 0.5 μm). (d) Calculated electromagnetic fields distribution in the hybrid plasmonic microcavity. 5 nm SiO₂ layer is marked by the dashed lines. Hybrid surface plasmonic modes arise along the cavity interface and induce strong localized field.

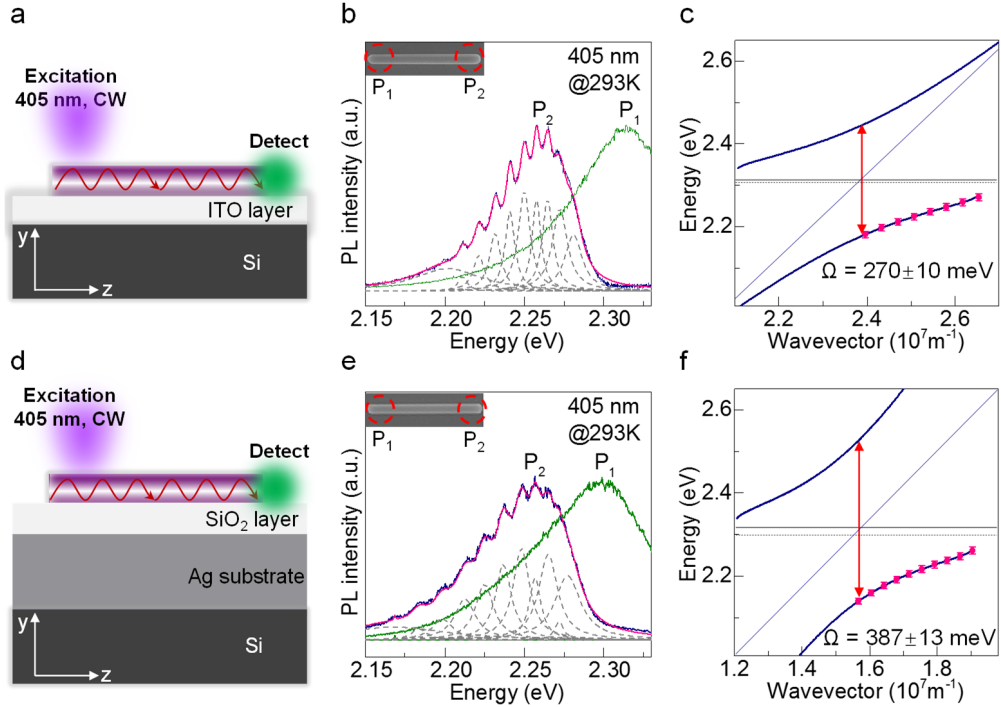

**Figure 2. Room Temperature Exciton-Polaritons in MAPbBr$_3$ NWs Sitting on Glass and SiO$_2$/Ag Substrates.** (a) and (d) Schematic of home-built remote excitation-collection confocal system for probing polaritons in NWs on glass (a) and 5 nm SiO$_2$/Ag substrates (d). (b) and (e) Photoluminescence spectra for NWs on glass (b) and 5 nm SiO$_2$/Ag substrates (e) detected at P$_1$ (green solid curve, in-situ excitation) and P$_2$ (navy solid curve, remote excitation). The MAPbBr$_3$ NWs is ~ 550±30 nm in width and 8.4±0.5 μm in length. (c) and (f) Energy-wavevector (*E-k*) dispersion in the z-direction for the NWs shown in (b) and (e). The horizontal solid and dashed lines indicate the longitudinal resonance energy $E_L$ and the transverse resonance energy $E_T$, respectively.

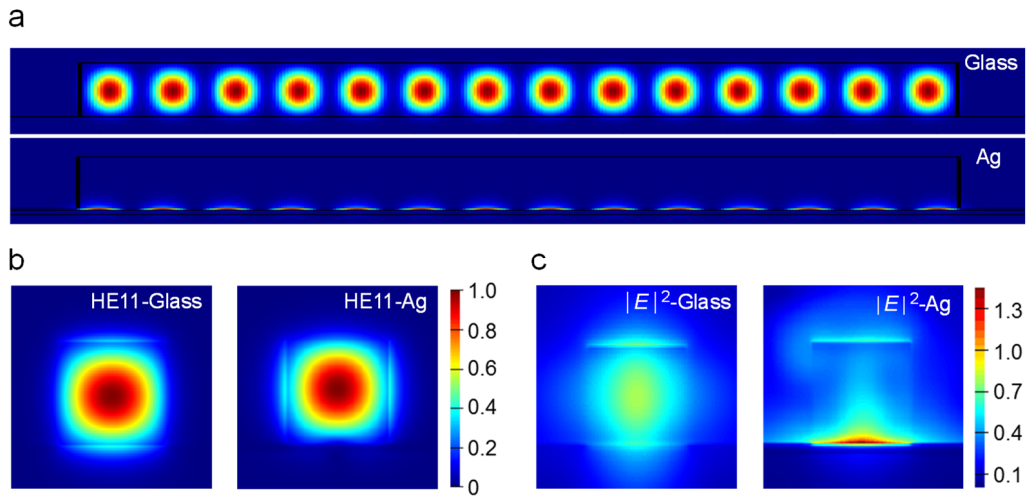

**Figure 3. Hybrid plasmonic/photonic modes on SiO$_2$/Ag substrates vs photonic modes on glass.** (a) Two-dimensional projections of the obtained three-dimensional mode distribution for the HE11 confined mode on glass and plasmonic mode on SiO$_2$/Ag substrate in a 550 nm wide and 8.4 μm long nanowire at λ = 550 nm. (b) Cross sectional plots of the normalized HE11 modes for nanowires with width of 550 nm on glass and 5 nm SiO$_2$/Ag substrates. (c) Electric field distributions in NWs with width of 550 nm on glass and 5 nm SiO$_2$/Ag substrates.

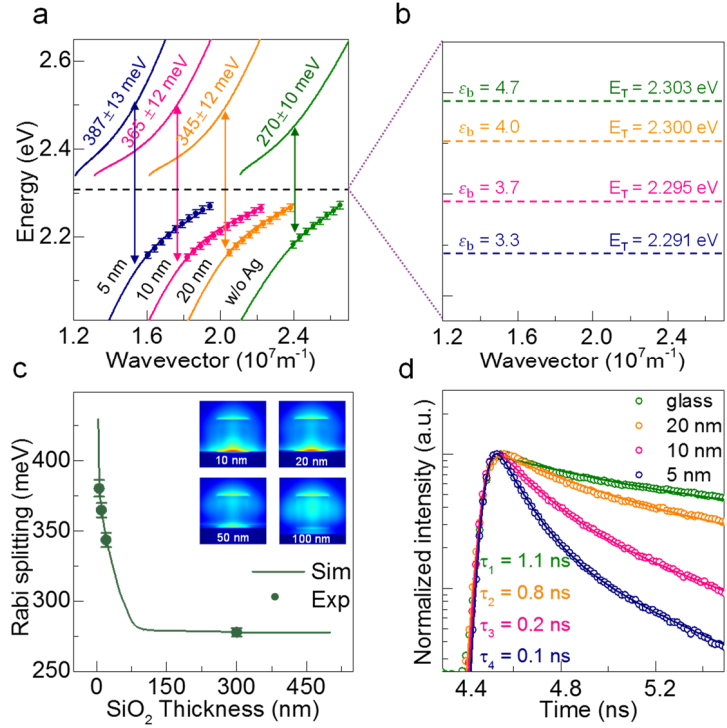

**Figure 4. Silica Thickness Dependence of Exciton-Photon Coupling in Hybrid Plasmonic Microcavity.** (a) *E-k* dispersions of NWs on glass and $SiO_2$/Ag substrates with different thicknesses of $SiO_2$. The nanowires' width and length are ~ 550±30 nm and ~ 8.4±0.5 μm, respectively. Rabi splitting and silica thickness are indicated along with each dispersion curve. (b) Background dielectric constant $\varepsilon_b$ and the transverse resonance energy $E_T$ for NWs on glass and $SiO_2$/Ag substrates. (c) Theoretical results (curves) and experimental value (points) of Rabi splitting energy versus the thickness of $SiO_2$. The insets: electromagnetic fields calculations of $MAPbBr_3$ NW microcavities with different thicknesses of $SiO_2$. (d) Time resolved PL spectra of $MAPbBr_3$ NWs sitting on glass and $SiO_2$/Ag substrates with $SiO_2$ of 5, 10, 20 nm, indicating a decreasing lifetime of 1.1 ns, 0.8 ns, 0.2 ns and 0.1 ns with the decreased thickness of silica. The PL lifetimes of NWs on the four substrates are indicated with same plotting color of experiment time decay curves.

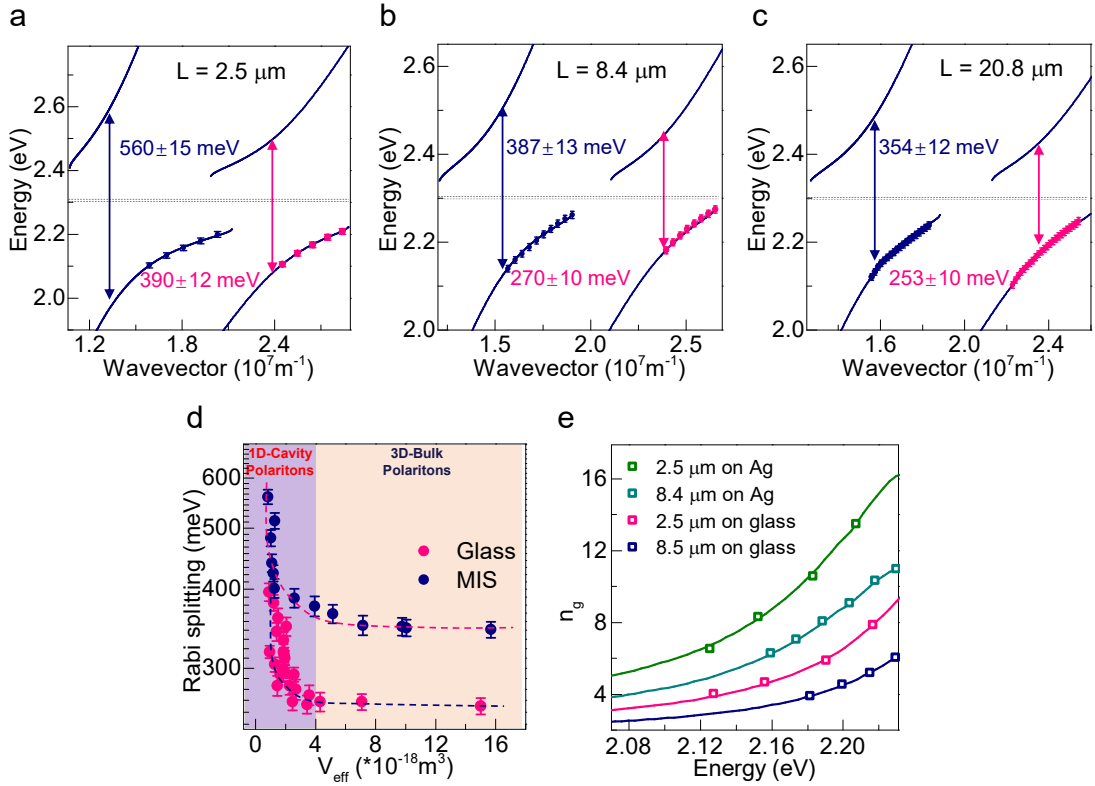

**Figure 5. Nanowire Size Dependence of Light-Matter Coupling.** (a)-(c) *E-k* dispersions in the *z*-direction of NWs on 5 nm SiO$_2$/Ag substrate (navy data points) and glass substrate (pink data points), respectively. The NW length is 2.5±0.5 μm (a), 8.4±0.5 μm (b), 20.8±0.5 μm (c) with comparable width of ~ 550 nm±30 nm. The upper and lower dash lines indicate transverse resonance energy ($E_T$ = 2.303 eV) for NWs on glass substrate and 5 nm SiO$_2$/Ag substrate ($E_T$ = 2.291 eV), respectively. (d) Effective mode volume dependence of Rabi splitting energy extracted out from the *E-k* dispersions for different NW microcavities on glass substrate and 5 nm SiO$_2$/Ag substrate with widths ranging from ~ 400-600 nm and lengths ranging from ~ 2.5-21 μm. The dashed curves: simulated results. (e) The group refractive index of NWs on glass and 5 nm SiO$_2$/Ag substrates. The NWs lengths: 2.5 μm (olive curve) and 8.4 μm (cyan curve) on 5 nm SiO$_2$/Ag substrate; 2.5 μm (pink curve) and 8.5 μm (navy curve) on glass substrate. The width is ~ 550±30 nm. The solid curves are calculated from the *E-k* dispersions by the formula, $n_g = (dE/dk)_{vacuum}/(dE/dk)_{cavity}$. The data points are obtained by calculating the group index using *E* and *k* of each oscillation peak in emission spectroscopy.